\documentclass[fleqn,12pt]{article}

\newcommand{\AmS}{{\protect\the\textfont2
  A\kern-.1667em\lower.5ex\hbox{M}\kern-.125emS}}

\title{Antineutrino Background from Spent Fuel Storage  in  sensitive  Searches  for  $\theta_{13}$ \\ at Reactors}

\author{V. Kopeikin, L. Mikaelyan, V. Sinev \\
\\
Russian Research Center "Kurchatov Institute"}
 
\begin{document}

\date{}
\maketitle


\begin{abstract}
Sensitive searches for antineutrino oscillations in atmospheric mass parameter region much discussed in recent years are based on accurate comparison of the inverse beta decay positron spectra measured in two (or more) detectors, far and near, stationed e.g. at $\sim$1000 m and $\sim$100\,m from the reactor(s). We show that antineutrinos emitted from the stored irradiated fuel can differently distort the soft part of positron spectra measured in the far and near detector and thus mimic (or hide) the oscillation signal
\end{abstract}

\section*{Introduction}

The $\sim$1 km baseline reactor experiment CHOOZ [1] searched for electron antineutrino energy dependent disappearance in the atmospheric mass parameters region using one detector. No oscillation have been found:

\begin{equation}
\sin^2{2\theta}_{CHOOZ} \le 0.20 \;(90\% \;{\rm CL \;for} \;\Delta m^2 = 2.0 \times 10^{-3}\;{\rm eV^2}).
\end{equation}

In the three active neutrino mixing scheme with normal neutrino mass hierarchy oscillation parameter $\sin^2{2\theta}_{CHOOZ}$ is expressed trough the contribution \ of \ the \ mass-3 \ eigenstate \ to \ the \ electron \ neutrino \ flavor \ state \\ $U_{e3} = \sin \theta_{13}$:

\begin{equation}
\sin^2{2\theta}_{CHOOZ} = \sin^2{2\theta}_{13} = 4 |U_{e3}|^2 {(1-|U_{e3}|^2)}. 
\end{equation}

Reactor antineutrinos are detected via the inverse beta-decay reaction

\begin{equation}
\qquad \qquad \qquad \qquad \bar{{\nu}_e}+p \rightarrow n + e^{+}
\end{equation}

The visible positron energy $E_e$ in scintillation spectrometer is related to the $\bar{{\nu}_e}$ energy $E$ as

\begin{equation}
E_{e} = E - 1.80 + E_{annihil} \approx E - 0.8 \ (\rm MeV).
\end{equation}

\section{CHOOZ data analysis}

The CHOOZ ratio $R_{meas/calc}$ of measured to calculated for no oscillation case detection rates was found to be 

\begin{equation}
R_{meas/calc} = \rm 1.01 \pm 2.8 \% \ (stat) \pm 2.7 \% \ (sist).
\end{equation}

Ratio of measured to expected no oscillation positron spectra $R_{shape}$

\begin{eqnarray}
R_{shape} = \frac {S(E_e)_{meas}}{S(E_e)_{calc}} = const \cdot [1 - \sin^2{2\theta_{13}}\cdot \sin^2\frac{1.27 \ \Delta m^2\ {L}}{E}] \quad
\end{eqnarray}
was analyzed and combined shape + rate analysis performed ($L$ is the reactor-detector distance).

The CHOOZ oscillation limit (1) was obtained using absolute method of analysis. All available experimental information was compared to expected no-oscillation values. Thus the results directly depend on correct determination of reactor power, absolute value of $\bar{{\nu}_e}$ energy spectrum, nuclear fuel burn up effects, absolute value of spectrometer response characteristics etc. It is generally agreed that absolute method cannot provide significant improvements in searches for smaller $\theta_{13}$.

\section{Towards higher sensitivity}

Soon after the CHOOZ oscillation experiment was completed it was understood that the sensitivity of searches for reactor neutrino oscillation parameters could be significantly increased if one reactor $-$ two identical detector layout is used [2, 3]. The near detector in this scheme is monitoring absolute flux and spectrum of $\bar{{\nu}_e}$, while the far detector looks for deformation of this spectrum due to oscillation survival factor $P (\bar{{\nu}_e} \to \bar{{\nu}_e})$:

\begin{equation}
P (\bar{{\nu}_e} \to \bar{{\nu}_e}) = 1 - \sin^2{2\theta_{13}}\cdot \sin^2\frac{1.27 \ \Delta m^2 \ L_{Far}}{E} 
\end{equation}

International neutrino community in 2002$-$2003 yy has analyzed perspectives of sensitive searches for $\theta_{13}$. A number of relevant new reactor experiments have been considered including two reactor $-$ two detector experiment at CHOOZ (Double CHOOZ), experiment in China (Daya Bay), multi reactor $-$ multi detector experiment in Japan (the KASKA Project), experiments in US and in Brazil \dots The results of this activity are summarized in White Paper [4].
Quite recently Letters of Intent for Double CHOOZ [5] and for US participation in Double CHOOZ [6] have been issued.

Main results of the analysis \ [2,3,4,5,6] \ can be formulated as following:

It is possible to reach the sensitivity of

\begin{equation}
\sin^2{2\theta}_{13} \le 0.02-0.03 \;(90\% \;{\rm CL \;for} \;\Delta m^2 = 2.0 \times 10^{-3}\;{\rm eV^2}), 
\end{equation}
if a number of prescriptions are followed and among them:

\begin{itemize}
\item Suppress the background/neutrino signal ratios in far and near detectors to better then 1:100 (since with 2 and more reactors no background measurements are expected to be possible).
\item Keep total systematic uncertainty below 0.6\%.
\item Increase the statistics of $\bar{{\nu}_e}$ events in far detector by a factor $\sim$30 (and more) relative to the CHOOZ experiment [1].
\end{itemize}

\section{Antineutrinos \ emitted \ from \ the \ stored \ irradiated fuel}

Antineutrinos emitted from the stored spent nuclear fuel (SNF), could spoil the first and the second conditions (see above) and, if not properly taken into account, introduce unwonted distortion in measured positron spectra and thus influence the interpretation of data analysis.

Some of the fission fragments, which emit $\bar{{\nu}_e}$ of energy higher than 1.80 MeV are in equilibrium with preceding long lived fission products and continue to emit $\bar{{\nu}_e}$ after the reactor is shut down at the end of the operational run [7]. Among mentioned fragments e.g. are:

$$
^{106}{\rm Ru} \ ( T_{1/2} = 372 \ {\rm d}) \to  ^{106}\!\!{\rm Rh} \ ( T_{1/2} = 30 \ {\rm s}, E_{max} = 3.54 \ {\rm MeV}) \quad
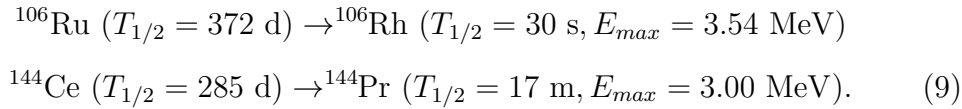
$$

\begin{equation}
^{144}{\rm Ce} \ ( T_{1/2} = 285 \ {\rm d}) \to  ^{144}\!\!{\rm Pr} \ ( T_{1/2} = 17 \ {\rm m}, E_{max} = 3.00 \ {\rm MeV}). \qquad
\end{equation}

Some contribution comes also from $^{90}{\rm Y} \ (T_{1/2} = 64 \ {\rm h}, E_{max} = 2.28 \ {\rm MeV})$, which is in equilibrium with its very long-lived predecessor $^{90}{\rm Sr} \ (T_{1/2} = 28.6 \ {\rm yr})$.

Regularly (annually) after the end of the operational run some part (often it is 1/3) of the irradiated fuel ("the spent nuclear fuel") is removed from the reactor and placed in a water pool not far from the reactor. Water provides cooling and shielding. After some period of time (may be $\sim$5 yr) each portion of the SNF is transported from the pool to the dry storage away from the nuclear power plant. Thus not far from the reactor is formed and regularly supported an additional, relatively small but not negligible, source of SNF $\bar{{\nu}_e}$ (see Fig. 1) capable to produce reaction (3) in the antineutrino detectors.

\section{Imaginary oscillation experiment with SNF $\bar{{\nu}_e}$ source}

Clearly \ expected \ effect \ due \ to \ source \ of \ SNF \ $\bar{{\nu}_e}$ \ critically \ depends \ on reactors $-$ detectors SNF pool(s) locations. As an example here we consider geometry presented in Fig. 2. It can be shown that in the case considered the ratio $R_{F/N}$ of the positron spectra measured in the far and near detectors can be presented as:

\begin{eqnarray}
R_{F/N} \approx {\rm C} \cdot [1 - \sin^2{2\theta_{13}}\cdot \sin^2 \frac{1.27 \ \Delta m^2 \ L_{Far}}{E} + \nonumber\\ + {\frac{S_{SNF}}{S_R}}(1-\frac {L^2_{Near}}{ (L_{Near} - L_{Pool})^2})]
\end{eqnarray}

Here $E$ is energy of the incoming $\bar{{\nu}_e}$ related to positron visible energy through Eq. (4). Distances $L$ between $\bar{{\nu}_e}$ sources and detectors are shown in Fig. 2. The constant C in (10) is (assumed to be) known within accepted systematic uncertainty of 0.6\%. The small term $(S_{SNF}/S_R) [1- (L^2_{Near} / (L_{Near} - L_{Pool})^2 )]$ represents correction stemming from spent fuel pool antineutrinos, $S_R$ is the reactor $\bar{{\nu}_e}$ energy spectrum; the ratio $S_{SNF}/S_R$ is shown in Fig. 1.

With distances shown in Fig. 2 the spent fuel distortion is negative, i.e. has the same sign as oscillation effect, which is sought for (see Fig. 3).

\section*{Conclusions}
 
Antineutrinos from spent fuel stored in pools near the reactor can undesirably distort the ratio of measured far/near positron spectra in planned $\theta_{13}$ experiments. The spent fuel effect ought be calculated and corrected for. This would require knowledge of the relevant fission products concentrations in the spent fuel and storage schedule in the water pool(s).

\section*{Acknowledgments }

This work is supported by the Russian Foundation of Basic Research and was also funded with a grant in support of leading scientific schools.

\appendix


\begin{thebibliography}{9}
\bibitem {Ap} M. Appolonio $et \ al.$, (CHOOZ Collab., Phys. Lett. B {\bf 466}, 415 (1999).
\bibitem {Mi} L. Mikaelyan, V. Sinev, Phys. Atom. Nucl. {\bf 63}, 1002 (2000); \\ arXiv:hep-ex/9908047.
\bibitem {Mik} L. Mikaelyan, Nucl. Phys. B (Proc.Suppl.) {\bf 87}, 284 (2000); \\ arXiv:hep-ex/9910042. \\
L. Mikaelyan, Nucl. Phys. B (Proc.Suppl.) {\bf 91},120 (2001); \\ arXiv:hep-ex/0008046.
\bibitem{Wh} White  Paper Report on Using Nuclear Reactors to Search for a value of $\theta_{13}$, http://www.hep.anl.gov/minos/reactor13/white.html.
\bibitem{Ar} F. Ardellier $et \ al.$, arXiv:hep-ex/0405032.
\bibitem{Be}S. Berridge $et \ al.$, arXiv:hep-ex/0410081. 
\bibitem{Ko} V. Kopeikin, L. Mikaelyan, V. Sinev, Phys. Atom. Nucl. {\bf 64}, 849 (2001); \\ arXiv:hep-ph/0110290.
\bibitem{Kop} V. Kopeikin, Phys. Atom. Nucl. {\bf 66}, 472 (2003); \\ arXiv:hep-ph/0110030.

\end{thebibliography}
\end{document}